\documentclass[aps,twocolumn,epsf,floats,prl,reprint,nofootinbib]{revtex4-1}
\usepackage{graphics,graphicx,epsfig}
\usepackage{amssymb,color}
\usepackage{epsf,epstopdf,wrapfig}
\usepackage{amsmath}
\usepackage{multirow}
\usepackage{color,graphicx}
\usepackage{amsmath}

\begin{document}

\title{On why a few points suffice to describe spatiotemporal large-scale brain dynamics}

\author{Ignacio Cifre$^{1,2}$}
\author{Mahdi Zarepour$^{4,5}$}
\author{Silvina G Horovitz$^{2,3}$}
 \author{Sergio Cannas $^{4,5}$}
 \author{Dante R Chialvo$^{2,5}$}

\affiliation{$^1$ Facultat de Psicologia, Ci\`encies de l'educaci\'o  i de l'Esport, Blanquerna, Universitat Ramon Llull, Barcelona, Spain}
\affiliation{$^2$ Center for Complex Systems $\&$ Brain Sciences (CEMSC$^3$), Universidad Nacional de San Mart\'in, 25 de Mayo 1169, San Mart\'in, (1650), Buenos Aires, Argentina.}

\affiliation{$^3$ National Institute of Neurological Disorders and Stroke, National Institutes of Health, Bethesda, MD, USA}
\affiliation{$^4$ Instituto de F\'isica Enrique Gaviola (IFEG), Facultad de Matem\'atica, Astronom\'ia y F\'isica, Universidad Nacional de C\'ordoba, Ciudad Universitaria, (5000), C\'ordoba, Argentina}
\affiliation{$^5$ Consejo Nacional de Investigaciones Cient\'ificas y Tecnol\'ogicas (CONICET), Godoy Cruz 2290, Buenos Aires, Argentina}

\date{\today}
\begin{abstract}
An heuristic signal processing scheme recently introduced shows how brain signals can be efficiently represented by a sparse spatiotemporal point process. The approach has been validated already for different relevant conditions demonstrating that preserves and compress a surprisingly large fraction of the signal information.  In this paper the conditions for such compression to succeed are investigated as well as the underlying reasons for such good performance.  The results show that the key lies in the correlation properties of the time series under consideration. It is found that signals with long range correlations are particularly suitable for this type of compression, where inflection points contain most of the information.  Since this type of correlation is ubiquitous in signals trough out nature including music, weather patterns, biological signals, etc., we expect that this type of approach to be an useful tool for their analysis.  
 \end{abstract}
 
%\keywords{}
\maketitle

In the analysis of complex spatiotemporal patterns, such as large scale brain dynamics, an important challenge 
is the adequate coarse graining of the data. In the case of brain imaging the dataset is composed of several thousand time series, of the so called BOLD (``blood oxygenated level dependent'') signal, covering the entire brain. The usual question in this analysis revolves around the detection of burst of correlated activity  across certain regions, which requires extensive computations, in part due to the usually humongous size of the data sets. 

Recently it was uncovered\cite{T1,T2,T3,caballero,allan} that these type of problems can be efficiently analyzed using only the timings of the peak amplitude signal events, i.e., a point process (PP).  Subsequent work using similar approaches\cite{li,liu1,liu2,chen,jiang,Amico,Wu} further confirmed that the method entails a large compression of the original signals. Overall these findings not only suggest  a way to speed up computations, but most importantly highlight the need to clarify which aspects  or features of the brain imaging signals contain the most relevant information. 

The present work is dedicated to clarify the reasons underlying the effectiveness of this approach. The results show that the key lies in the correlation properties of the time series under consideration. In synthesis, it is found that signals with long range correlations are particularly suitable for this type of compression, where inflection points contains most of the information.  The results applied as well to other signals from any origin as  long as their correlation features are similar.

%%%%FIGURE 1%%%%%%%%%%%%%%%%%%%%%
\begin{figure} 
\begin{center}
\includegraphics[width=.4\textwidth]{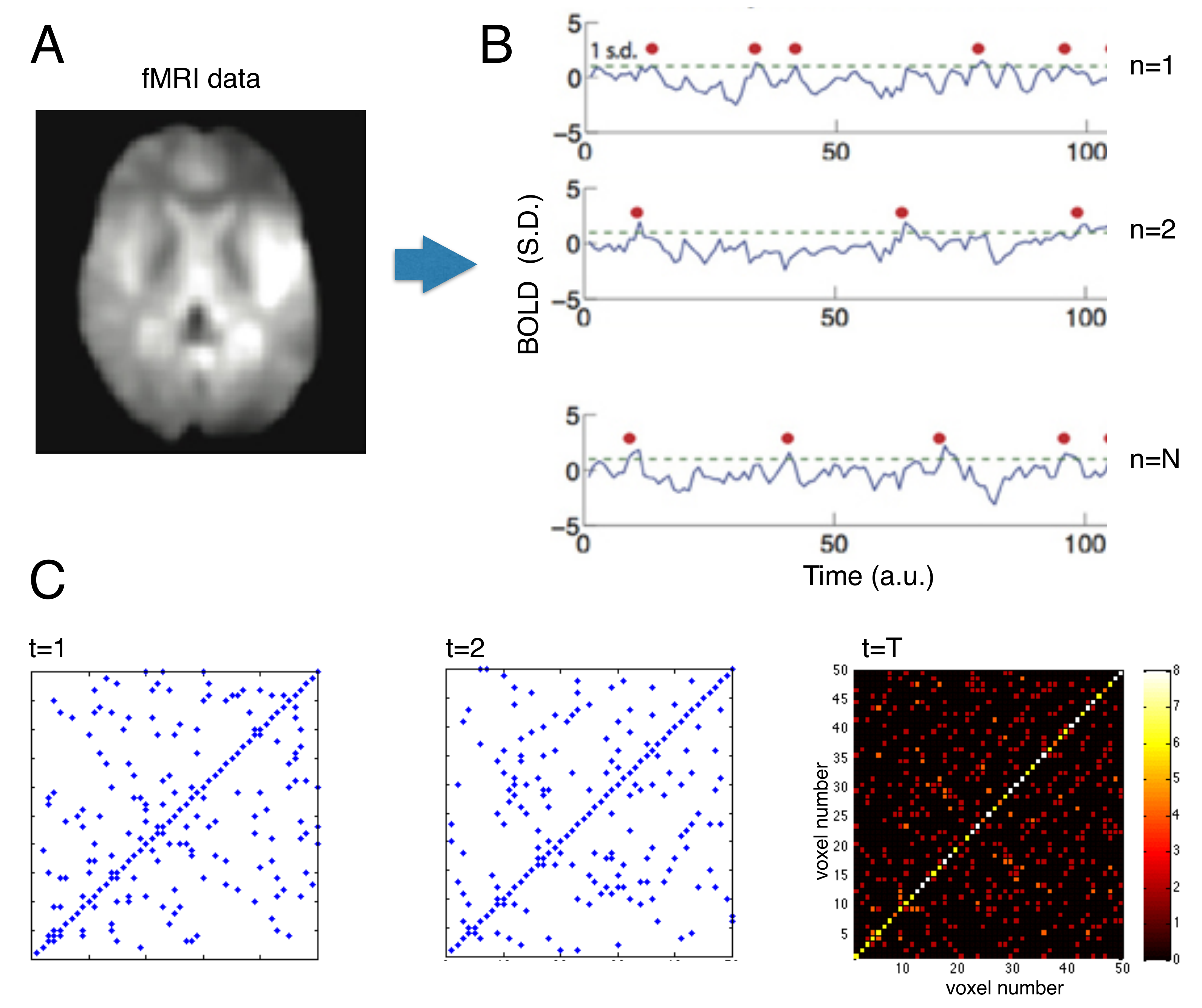}
\caption{The basic process used to extract the point process. The traces on panel B are examples of time series of fMRI BOLD signals at three brain locations (so called ``voxels''). Time points are selected at the upward threshold crossings or the peaks of the signal (filled circles). The temporal co-occurrence of these points defines co-activation matrices for different length of time (graphs in C ) which can be  further averaged to estimate the correlation matrix for the entire time T of the system under study. }
\label{cartoon}
\end{center}
\end{figure}

Figure 1 summarizes the basic process that has been used in\cite{T1,T2,T3} to define the point process in brain signals.  The data consists in time series representing the activity of one of many thousands small brain regions, recorded from the brain using functional magnetic resonance imaging (fMRI). This imaging technique measures in each small region  a ``blood oxygenated level dependent'' signal (i.e., ``BOLD''), that is an estimation of the blood' saturation of oxygen, which itself  is proportional to the local neuronal activity.   As shown in the figure, time points are selected at the upward threshold (here at unity) crossings of the signal (filled circles). The  point process  can be constructed also by selecting the local peaks of the BOLD time series. The temporal co-occurrence of the points defines the co-activation matrix (bottom graphs) which can be  further averaged to estimate the correlation matrix of the system under study. 

 \begin{figure}[h]
\centering
\includegraphics[width=.45\textwidth]{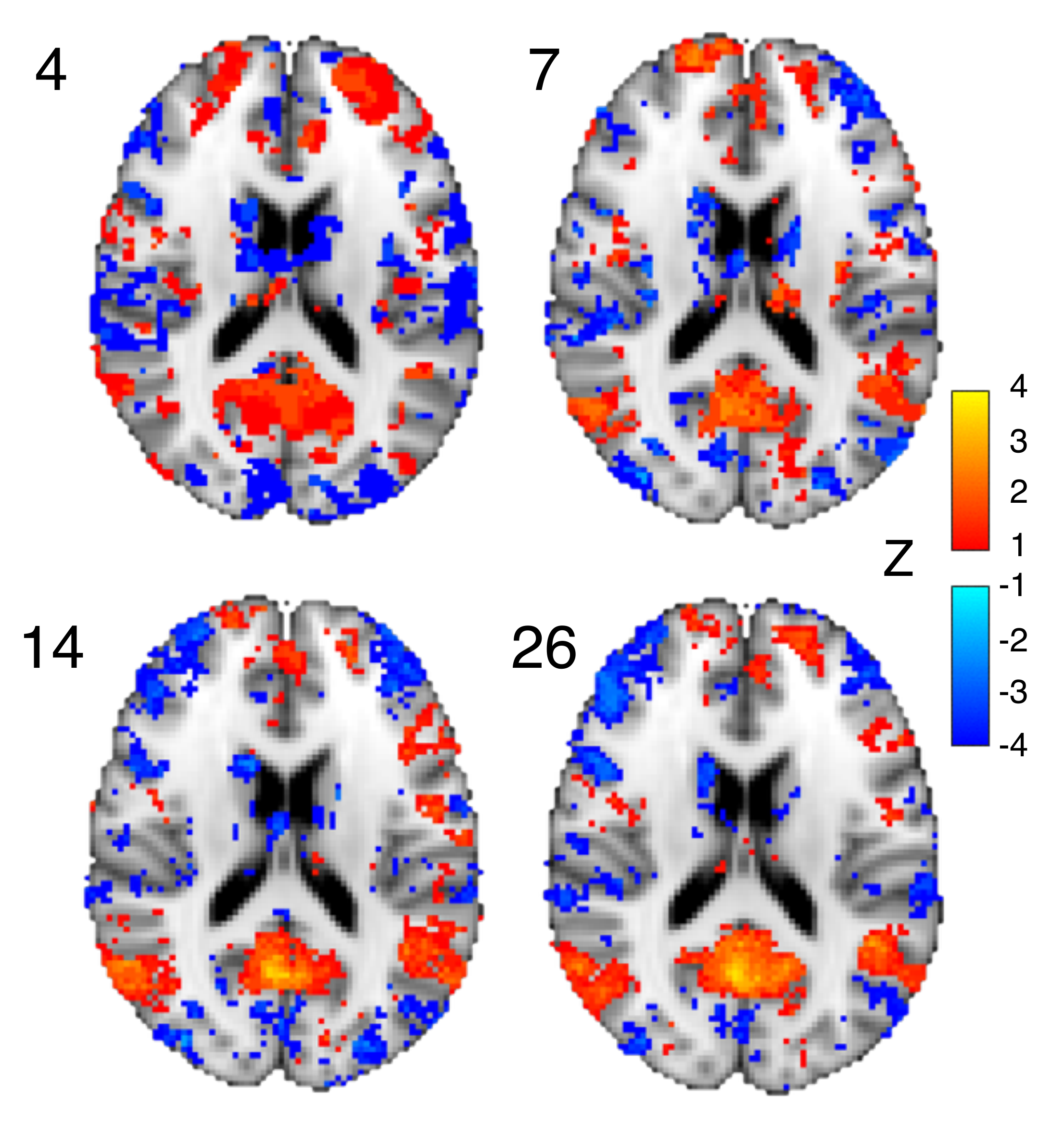}
\caption{An example of the performance of the PP as a function of the number of points used. The ``heat map'' images  represent the co-activation rates (in units of standard deviation, z-scores $Z$ ) of each voxel respect to the seed located at MNI coordinates x=4, y=-60, z=18. Different panels correspond to the results obtained using increasing number of points: 4, 7, 14 and 27, respectively.  Note that a few points already suffice to identify well defined clusters that are 1-4 standard deviations away from chance co-activations. Red/ blue colors label positive/negative points co-activations, corresponding to positive/negative correlations.}
\end{figure}
 
It has been established already, in different circumstances\cite{T1,T2,T3}, that the co-activation matrix obtained with the PP methods is very similar to the correlation matrix computed from the full  (i.e., continuous)  BOLD signal. Since this implies a large compression, the question is { \it why a few points are enough to compute  results similar to those obtained with the full signal}. Figure 2 shows an example constructed from BOLD time series from an experiment in which the subject is resting \cite{T3}. The results demonstrate  that as few as 4 points are already sufficient to define clusters of co-activation, as demonstrated previously in \cite{T1,T2,T3}. In addition, the results here show how de-activations (i.e., blueish colors) are also evidenced by the PP approach. 

A simple visual inspection  of the BOLD traces reveals that the type of signals we are dealing with are temporally correlated. This is very well known, the neuronal activity is temporally and spatially correlated, and furthermore the activity is convoluted by the hemodynamic transfer function which in itself introduce additional temporal correlations. Therefore, for any time series with that properties, it seems natural to think that the most informative points  are those in which its derivative changes sign. The rest of the points are redundant, since they can be predicted, up to a degree, by a linear estimator. 
%%%%FIGURE3%%%%%%%%%%%%%%%%%%%%%
\begin{figure}[h]
\centering
\includegraphics[width=.47\textwidth]{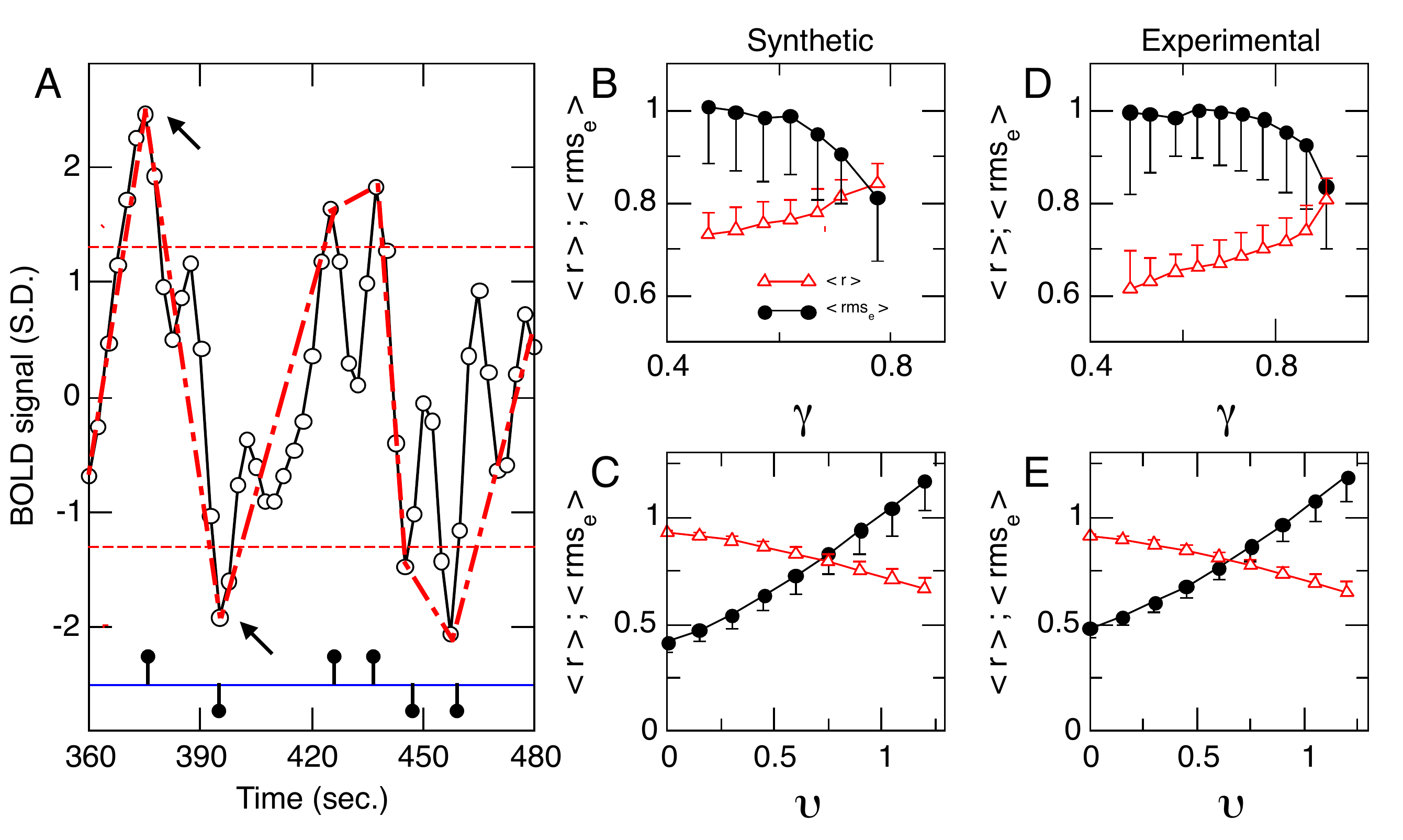}
\caption{Why it works: The  trace in Panel A  is an example of a two minutes-long recording of a BOLD brain signal during rest. The point process is defined by the peaks and valleys,  encompassed only by the six dots depicted in the bottom trace out of the 120 points of the time series. A piece-wise linear time series (dashed lines) is constructed by joining the peaks and valleys (two indicated by arrows) larger than a given threshold, here denoted by horizontal dotted lines.  Panels D and E correspond to the computed similarity between the BOLD signals and the piece-wise linear signals (evaluated by the correlation $\langle r \rangle$ and error $rms_e$ values) for different autocorrelation $\gamma$ and  threshold $\nu$ values. Panels B and C correspond to similar calculations for synthetic time series.}
\end{figure}
%%%%%%%%%%%%%%%%%%%%%%%%%%%%%% 

This  is illustrated in Figure 3 using as an example two minutes of  BOLD recording (normalized by its standard deviation (S.D)).  After setting a threshold $\nu$ the inflection points larger than a given $\nu$ value are identified. These points constitutes the marked point process  in question. Now we ask how much of the raw signal is left out if these points are used to extrapolate a piece-wise linear time series. To answer that we analyze BOLD time series from the brain of a subject during an experiment in which fMRI data is collected at rest\cite{T2}. We proceed to compute the linear correlation between the two time series, the raw and the piece-wise linear one. In panels B and C are shown the results for different values of threshold $\nu$ (in units of S.D.) as well as for the correlation of the time series, estimated by the value of the first autocorrelation coefficient $\gamma$.
Panel D shows that as the  BOLD signal' autocorrelation increases the similarity between the piece-wise linear and the raw signals increases, evaluated in two ways: by the $rms_e$ error and the linear correlation $\langle r \rangle$ between both time series. As expected, the  raising of the threshold $\nu$ from zero (i.e, less information from the signal is considered) it is followed by a monotonic increase of the  $rms_e$ and a decrease of the $\langle r \rangle$ values (see Panel E).

According with the present hypothesis, the functional dependences shown by the  BOLD signals in Panel D and E shall be replicated by using  synthetic signals with similar autocorrelation properties. For that we generate artificial time series with autocorrelation values identical to those of the BOLD signals using the routine {\verb f_alpha_gaussian.m } from MATLAB.  Panels B and C show that the dependence with the threshold $\nu$ and $\gamma$ exhibit very similar behavior.
The results show that the key lies in the correlation properties of the time series under consideration. In synthesis, it is found that signals with long range correlations are particularly suitable for this type of compression, where inflection points contains most of the information. The results shall apply as well to other signals from any origin as  long as their autocorrelation features are similar.

In summary, the success and the merits of the PP approach to represent spatiotemporal dynamics are related to a very trivial fact: in the case of autocorrelated signals the only informative points are those with zero derivative (inflection points); remaining ones are more or less straight lines which can be in principle, and for certain applications, ignored. Applications of these ideas to a diversity of fields should be expected.

\end{document}